\documentclass[12pt]{article}
\usepackage{graphicx}
\def\asec{$^{\prime\prime}~$}

\def\Msun{~M$_{\odot}~$}

\def\um{~$\mu$m}
\def\ul{\underline}

\begin{document}

\renewcommand{\theenumi}{\alph{enumi}}
\renewcommand{\labelenumi}{\theenumi.}

\large
\centerline{\bf First Steps in Direct Imaging of Planetary Systems Like our Own:}
\centerline{\bf The Science Potential of 2-m Class Optical Space Telescopes}
\normalsize
\vskip 0.1 in
\centerline{Karl Stapelfeldt\footnote{Mail Stop 183-900, JPL, 4800 Oak Grove Drive, Pasadena CA 91109, 818-354-2640, krs@exoplanet.jpl.nasa.gov},
John Trauger, Wesley Traub (NASA-JPL)}
\centerline{Mark Clampin, William Oegerle, Jennifer Wiseman (NASA-GSFC)}
\centerline{Olivier Guyon (Subaru Telescope)}

\vskip 0.4 in

\leftline{\bf 1.~~~The Exoplanet Science Landscape in 2015}

A planetary system consists of gas giant planets, rocky terrestrial
planets, and belts of small bodies which generate debris particles.
Ongoing research and upcoming instrumental developments promise to
significantly advance our knowledge of these three exoplanetary system
components in the coming decade.  Radial velocity surveys have already
found a planet with half the mass of Neptune around an M star (Rivera
et al. 2005).  Around Sun-like stars, they will probe for Neptune-mass
planets orbiting within 1 AU and Jupiter-mass planets orbiting as distant
as 7 AU.  For the very low mass stars, they may achieve detections of
objects as small as a few Earth masses at orbital distances of a few
tenths of an AU (Udry et al. 2007).  The orbital element distribution 
for the inner jovian planets of nearby stars should be well in hand 
by 2015.

Upcoming transit observations will reveal the frequency and
radii of close-in (a $<$ 1 AU)  rocky exoplanets by photometrically
monitoring large ensembles of solar-type stars.  The COROT mission
will find examples of large terrestrial planets with orbital
periods $<$ 2 months by 2009, while Kepler will find Earth mass
planets in 1 AU orbits by 2013.  Near-infrared studies have the
potential to do the same for M stars (where the small stellar
size maximizes the transit depth).  Numerous hot Jupiters will
be found, including some by their integrated reflected light.
Spectroscopic measurements made during transit and secondary
eclipse by Spitzer, the upgraded HST, and JWST will constrain
the albedos of these planets, and detect a few very high opacity
atmospheric species such as Na I and Ly $\alpha$ (Charbonneau et
al. 2007).  By 2015, transit work should have yielded a strong
statistical understanding of the inner	parts of extrasolar
planetary systems.

High contrast imaging detections of brown dwarf companions to
nearby stars are anticipated from large groundbased telescopes
deploying the next generation of adaptive optics systems, and
from space using JWST and perhaps Spitzer.  Contrasts approaching
10$^{-7}$ may be achievable from the ground in the near-infrared
(Beuzit et al. 2007), which would enable the detection of warm
(young/massive) gas giant planets at separations $>$ 0.2\asec with 10-m
class telescopes.  An appropriately designed 30-m telescope would
enable companion detections at even smaller inner working angles.
At 4.5 \um, JWST/NIRCam coronagraphy will be capable of detecting
companions at contrasts of 10$^{-6}$ at separations beyond 1.5\asec˝,
capturing objects like our own Jupiter in thermal emission
as companions to the nearest M stars.  The uncertain luminosity 
evolution of young giant planets clouds the picture somewhat (Marley 
et al. 2005), but it appears that the outer, massive planets orbiting 
nearby (d$<$ 20 pc), young (age $\sim<$ 1 Gyr), low-mass (M$<$ 0.5 \Msun) 
stars could be in view by 2015.

Imaging of protoplanetary disks will be revolutionized by ALMA,
which will be able to resolve dynamical structures driven
by protoplanets at angular resolutions approaching 0.01\asec.
For the more nearby debris disks, ALMA will be able to map
systems brighter than 1000 zodis at 0.1\asec
resolution\footnote{A 5$\sigma$ detection of the Fomalhaut debris 
ring can be achieved at 450 \um~in 6 hours at this resolution, according
to the online ALMA sensitivity calculator}; at 20 \um~JWST will 
resolve systems around nearby A stars with 0.3\asec˝resolution; 
and large ground telescopes with adaptive optics will push toward 
the 1000 zodi sensitivity level in the near-infrared.  A wealth 
of new data detailing the internal structure of bright circumstellar 
disks will be emerging in 2015, seeding a new theoretical understanding 
of disk structure, dynamics, and evolution.

While the advances described above will be remarkable scientific
milestones, they fall well short of the goal of obtaining images
and spectra of planetary systems like our own.	Transits will
detect inner terrestrial planets around distant stars,	but
spectroscopic characterization of them is highly unlikely.
High contrast imaging will detect and characterize warm giant
planets, but not cool objects at 10$^{-9}$ contrast like our own
Jupiter and Saturn in their orbits around a solar-type star.
Sharp images of dusty debris disks will be obtained, but only
those with optical depths $\sim>$1000 times that of our own asteroid
and Kuiper belts.  Radial velocity surveys have singled out
the nearby stars that host Jovian planets.  What is currently
missing from the 2015 exoplanetary science toolbox, and what we
now describe, is an observatory that can study photons from these
giant planets, and the terrestrial planets and exozodiacal dust
clouds that may accompany them, around nearby stars like the Sun.

\vskip 0.3 in
\leftline{\bf 2.~~~The Case for a Small Exoplanet Space Telescope Mission}

\noindent
\ul{The Programmatic Imperative}

NASA has performed significant design work on coronagraph and
interferometer concepts for a Terrestrial Planet Finder mission (Levine
et al. 2006; Lawson et al. 2007), and is now also considering occulter
concepts.  A common denominator of these design efforts is that they are
driven by the need to take spectra of terrestrial planets with V $\sim>$30.
Large telescopes in the 4-8-m class are required to achieve this at visible
wavelengths, or a constellation of 3-m telescopes in the thermal infrared.
Missions like these are central to a full exploration of neighboring
planetary systems; nothing would serve exoplanet science better than
beginning one of them as soon as possible.  However, all of these have the
scope of a flagship mission.  Without a major injection of new resources
into space astronomy, it will not be possible to realize a full-scale TPF
mission until many years after JWST has begun operations.  Our community
can either resign itself to waiting out JWST, or look for ways to achieve
significant new exoplanet science, sooner, through more modest projects.

A major opportunity for observational progress is offered by smaller
optical space telescopes.  These platforms have strong design heritage,
and provide the simplest approach to obtaining the high spatial that is
needed.  The science they would enable is the study of giant exoplanets 
around nearby stars in reflected light.  At V $\sim$27, these targets can 
be detected and characterized by low-resolution spectroscopy using a space 
telescope in the 2-m class, making the project small enough to proceed in 
parallel with JWST.  Such a mission is missing from NASA's current portfolio 
because of over-optimistic early assessments about the contrast achievable 
with groundbased adaptive optics (Angel 1994), and subsequent attempts to 
refit HST with an advanced coronagraph.  With 10$^{-8}$ now recognized as 
the useful contrast limit for groundbased AO (Angel et al. 2006; Guyon 2005b; 
Stapelfeldt 2006) and the end of Hubble instrument opportunities, the way 
is clear to consider dedicated space missions targeting exo-Jupiters and 
the means for their implementation.

\noindent
\ul {Science Reach of a 2m-class Optical Coronagraph}

\begin{enumerate}

\item {\it Studies of  Known Radial Velocity Planets}.  
Nine nearby stars host
radial velocity planets whose apastron distances project to angular
separations $\ge$ 0.25\asec.  These are cold objects presenting a contrast
of $\sim10^{-9}$ in the optical and near-IR, and thus are only accessible to a
space mission.  Existing ephemerides allow observations to be timed to
coincide with their maximum elongations.  A 2-m class coronagraph could
measure their colors, take spectra at R$\sim$40, and provide astrometric
measurements that will resolve the $\sin{i}$ ambiguity in their masses.
The spectra will allow CH$_4$ and NH$_3$ to be identified in the atmospheres,
and the depth of the uppermost cloud deck to be measured.  The planets'
albedo and presence of bright ring systems can be inferred using the
observed photometry and planet sizes estimated from their measured masses.
Multi-epoch imaging showing the planetary orbital motions will make a
powerful visual impression of the reality of exoplanetary systems in
the public mind.

\item {\it Discovery of New Giant Planet Companions}.  Radial velocity 
surveys are incomplete for orbital periods $>$ 8 years, for early F and hotter
stars lacking strong metallic lines in their spectra, for stars with
high chromospheric activity, and for planets in nearly face-on orbits.
Multi-epoch imaging with a 2-m class coronagraph has the potential
to discover new Jovian planets in 5-10 AU orbits around as many as 200
nearby stars.  There are 30 stars within 25 pc distance that host close-in
radial velocity planets, and which would be prime targets for an outer
planet search.  Spectral characterization of the detected objects would
also be carried out.

\item {\it Debris Disks and Exozodiacal Dust}.  Circumstellar dust disks are
maintained by ongoing collisions in belts of asteroidal and cometary
parent bodies.  In addition to revealing the location of these belts, 
debris disks act as a canvas on which unseen planets can impress dynamical 
signatures.  A 2-m class coronagraph provides the 0.1\asec spatial resolution 
needed to resolve the rings, warps, and asymmetries driven by planetary 
perturbations in these disks.  With contrast improved $\sim$ 1000 times over 
HST, it also will be sensitive enough to detect disks as tenuous as our own 
Kuiper Belt, enabling comparative studies of dust inventory and properties 
across stellar ages and spectral types.  

\item {\it Terrestrial Planets.}  For 5-10 of the brightest and nearest stars, 
the inner working angle of a 2-m class optical telescope may be good enough
to allow detection of Earth-sized planets within the habitable zone, and to 
spectrally characterize them using broadband photometry.  These results would 
provide a foretaste and motivation for a full-scale TPF mission to follow.

\end{enumerate}

\begin{figure}[t]
  \centering
  \begin{minipage}[b]{8 cm}
    \includegraphics[height=2.5in]{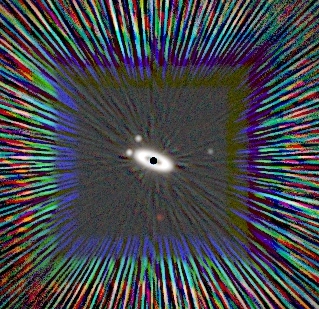}  
  \end{minipage}
  \begin{minipage}[b]{8 cm}
    \includegraphics[height=2.5in]{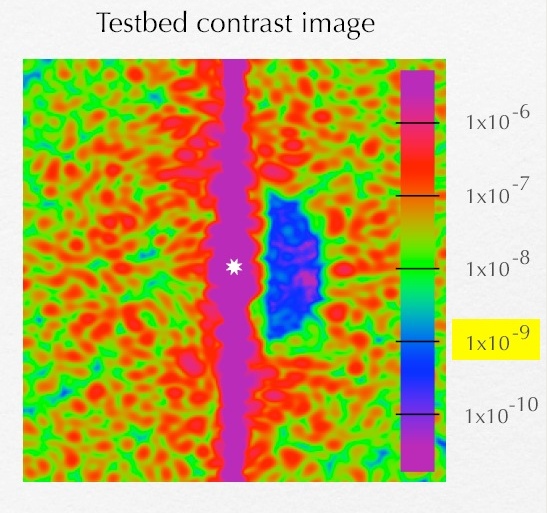}  
  \end{minipage}
  \caption{Left: Simulated coronagraphic image of a nearby planetary system, 
with Jovian planets and an inner 5 zodi dust ring as might be seen in VRI 
bandpasses with a 1.8-m telescope.  Right: Laboratory testbed image 
demonstrating the state of the art in optical wavefront control 
(Trauger \& Traub 2007)}

  \label{Labelname}
\end{figure}

\vskip 0.3 in
\leftline{\bf 3.~~~Technical Overview}

\noindent
\ul {Wavefront Control}  

Precision wavefront control is the key enabling technology for ultra-high
contrast imag-ing.  It is impractical to polish a large optic to the
smoothness required to directly detect exoplanets.  Instead, deformable
mirror(s) are used to actively correct the telescope mid-spatial frequency
wavefront to the very high accuracy of $\lambda$/1000.  In a space environment
these corrections are expected to be highly stable on timescales of days,
unlike the 10 ms timescales typical for groundbased adaptive optics.
Over the past six years this technique has been developed in the High
Contrast Imaging Testbed at NASA/JPL, to the point where contrasts
of 10$^{-9}$ are now achieved at radii of 4-10 $\lambda$/D in bandpasses of 2\%,
and a noise level from image to image of about 0.1 Earth or 0.01 Jupiter
(Trauger \& Traub 2007).  This level of wavefront correction is achieved
in a restricted in a ``dark hole'' region of the image field whose
spatial frequencies can be controlled by the finite deformable mirror.
Such mirrors are already available in a 64$\times$64 format that can support an
exoJupiter mission.  Work continues to refine the algorithms used to find
the optimal wavefront solution, and to reduce instrumental chromatism so
that broader bandpass corrections can be achieved.  The key point is
that the level of wavefront control needed to support 10$^{-9}$ image contrast
is close to being realized.  The testbed effort has been supported by
the TPF project, which has requirements 10 times more stringent than an
exoJupiter coronagraph mission.

\noindent \ul {Telescope Requirements}

The use of active wavefront corrections means that the telescope primary
does not need to be manufactured with extraordinary surface quality.
Instead, the key telescope requirement is mechanical stability.
The primary mirror must be protected from thermal stresses through
the use of proper materials, temperature regulation, and shielding
from outside thermal disturbances.  Primary-secondary alignment must be
carefully maintained, both to keep the pupil image on the deformable
mirror and to minimize focus drifts.  The inner working angle (IWA) that
the telescope provides to a coronagraphic instrument is determined by the
alignment stability that can be achieved.  For 10$^{-9}$ contrast, the usable
IWA is believed to be 3.5 $\lambda$/D for a large ($\sim$ 8-m) telescope, but
might be as small as 2.5 $\lambda$/D for a smaller, stiffer system.  While
it would be highly desirable to image planets at even smaller angular 
separations, the telescope alignment tolerances needed to enable operation 
at $<$ 2 $\lambda$/D are thought to be prohibitive.  Continued investigation
of the limiting IWA issues will be crucial for defining the sample of
terrestrial planets a 2-m class telescope might reach. 
The size of the telescope strongly affects the number of stars that
can be studied, as D$^3$.  A monolithic, unobscured (off-axis) telescope
design is preferable, as it maximizes throughput and substantially
simplifies diffraction control requirements.  Silver optical coatings
would be used to maximize system throughput between 0.5-1.0 \um, and
the reflectivity across the pupil should be uniform to within 0.1\% or
significant wavefront errors would be introduced.  A program to develop
a 1.8-m primary mirror meeting these requirements was begun by the TPF
project, but then suspended due to funding cuts.  The re-initiation
of this Technology Demonstration Mirror (TDM) effort would be highly
beneficial to the early realization of an exoJupiter coronagraph mission.

\noindent \ul {Diffraction Control}

A wealth of options are available to suppress stellar diffraction to
the levels needed to image exoJupiters.  A traditional Lyot coronagraph
has been employed in the JPL laboratory demonstrations, focusing on
graded masks that are insensitive to low-order telescope aberrations
(“band-limited masks”).  Masks made from HEBS glass have been used,
but wavelength-dependent phase shifts in this material may be limiting the
achievable bandwidth.  Promising alternative mask materials are under 
investigation.  A drawback of the Lyot coronagraphs is their relatively 
low throughputs $<$ 40\% and loss of resolution at the Lyot stop - even for 
unobscured apertures.  An alternative coronagaph design, dubbed Phase-Induced
Amplitude Apodization (or PIAA), offers much higher throughput and spatial
resolution close to that of the unobscured aperture (Guyon 2005a).  It is 
now undergoing initial laboratory demonstration.  Instead of blocking
the on-axis starlight, a third option is to null it out using a copy of
the input beam shifted in phase by 180 degrees.  Subsystem laboratory
testing has been done on the nulling coronagraph (Shao 2004), and
it will be the focus of an upcoming sounding rocket experiment.  A fourth
option is the finely tailored ``shaped pupil'' designs that can strongly
suppress diffraction along preferred directions in the image plane.
Finally, an external occulter formation-flying thousands of kilometers
from the telescope represents still another type of coronagraph.
This is likely to be a costly exoJupiter mission option, however,
because the expense of a second independent spacecraft would outweigh
the savings in telescope and wavefront control requirements.

\noindent \ul {Backend Instruments}

Characterization of exoJupiters in reflected light requires an imager
and spectrograph that operate over the wavelength range 0.5-0.95 \um.
Existing low-noise optical CCDs would be sufficient as the detectors,
and the required format would be relatively small (512 or 1024 square).
A spectral imaging instrument (Integral Field Unit) would be ideal, to
enable simultaneous detection, spectral characterization, and speckle
rejection.  Other configurations such as a staring camera with filter
wheel, dichroic separation into multiple simultaneous imaging channels,
a long-slit spectrograph, or the long-sought energy-discriminating
detectors could also be employed.  The optimal instrument approach
will strongly depend on the bandwidth that is achieved by the wavefront
correction system.  The science camera will also be used to carry out
the crucial function of wavefront sensing.

\noindent \ul {Spacecraft and Mission Design}

The spacecraft must provide a quiet and stable platform for the telescope
payload while handling routine functions such as power, data handling,
communications, and attitude control.  Pointing would be done through a
combination of coarse spacecraft body control, and fine steering of the
stellar target onto the occulting spot using internal instrument mirrors.
To minimize thermal disturbances to the telescope that would induce
alignment instability, an Earth-trailing or L2 orbit is preferred.
Studies to date suggest that a 2-m class payload can be injected
into these orbits using a Delta II class launcher.  If sent to L2,
the spacecraft would require an on-board propulsion system for orbit
maintenance.  A mission lifetime of 3-5 years is required to search for
outer planets with 12-20 year orbital periods; this gives the planets
enough time to move from inferior conjunction to maximum elongation,
where they will be bright and well-separated from their parent star.

\vskip 0.3 in
\leftline{\bf 4.~~~A New Programmatic Approach}

A 2-m class coronagraphic space telescope is an important first step toward
the goal of directly imaging and spectrally characterizing planetary
systems around nearby stars like the Sun.  It would produce important
science, set the stage for subsequent TPF missions, and is close to
technical readiness.  Unfortunately, NASA lacks a programmatic avenue
that would allow this class of mission to go forward.  The NASA Navigator
Program currently encompasses only groundbased efforts, SIM, and the
large TPF missions; no provision has been made for ``probe'' missions of
intermediate scale.  The community has tried to overcome this omission,
with several groups repeatedly proposing 1.2-1.8-m coronagraphic telescopes
to the NASA Discovery Program.  These have consistently been declined,
primarily on the basis of cost.  {\bf The situation is at an impasse: unless 
something is done, direct imaging of planetary systems like our own will 
be left waiting more than a decade for the coveted flagship missions, and 
unable to pursue important, more affordable missions that could be realized 
on a 5 year timescale.}  The NASA New Frontiers Program represents the mission
class (\$650 M) that could support the scope of a 1.5-2-m coronagraphic
telescope.  A specific exoplanet probe mission opportunity like this is 
urgently needed under the auspices of the Navigator Program.  In such a 
vehicle, a 2-m class coronagraph could be weighed against other exoplanet
probe mission concepts, with an open competition for the best instrumental 
approaches.  It is urgent that the Exoplanet Task Force consider this or 
other options that would enable near-term progress in the direct detection 
of nearby planetary systems like our own.

\vfil\eject
\leftline{\underline{References}}

\noindent
Angel, R.J.P. 1994, Nature 386 203 

\noindent
Angel, R.J.P, Codona, J., Hinz, P., Close, L. 2006, SPIE 6267 73

\noindent
Beuzit, J.-L., Mouillet, D., Oppenheimer, B.R., Monnier, J.D. 2007 in 
{\it Protostars and Planets V}, Univ. of Arizona Press, pp. 717-732

\noindent
Charbonneau, D., Brown, T., Burrows, A., Laughlin, G. 2007 in 
{\it Protostars and Planets V}, Univ. of Arizona Press, pp. 701-716

\noindent
Guyon, O. 2005a Ap.J. 622 744

\noindent
Guyon, O. 2005b Ap.J. 629 592

\noindent
Lawson, P.R., Lay, O.P, Johnston, K.J., Beichman, C.A. Eds. 2007, TPF-I SWG Report, 
JPL Publication 07-1

\noindent
Levine, M., Shaklan, S., Kasting, J. eds. 2006, TPF-C STDT Report,  JPL document D-34923

\noindent
Marley, M., Fortney, J., and Hubickyi, O. 2005 DPS 37 25.05

\noindent
Rivera, E.J. et al. 2005 Ap.J. 634 625

\noindent
Shao, M. et al. 2004 SPIE 5487 1296

\noindent
Stapelfeldt, K.R. 2006 in {\it The Scientific Requrements for Extremely Large Telescopes}, 
IAU 232 Proceedings, Cambridge Univ. Press, pp. 149-158.

\noindent
Trauger, J.T. and Traub, W.A. 2007 Nature April 12.

\noindent
Udry, S., Fischer, D., and Queloz, D. 2007 in {\it Protostars and Planets V}, Univ. of 
Arizona Press, pp. 685-700

\end{document}